\documentclass[a4paper]{PoS}
\usepackage{amsmath}
\usepackage{amssymb}
\usepackage{dsfont}
\usepackage{booktabs}
\usepackage{extarrows}
\usepackage{iftex}
\usepackage{enumitem}
\usepackage{tikz}

\ifPDFTeX
	\usepackage[utf8]{inputenc}
	\usepackage[T1]{fontenc}
\else
	\usepackage{fontspec}
	\usepackage{polyglossia}
	\setmainlanguage{english}
\fi
\usepackage{placeins}
\makeatletter
\AtBeginDocument{%
  \expandafter\renewcommand\expandafter\subsection\expandafter
    {\expandafter\@fb@secFB\subsection}%
  \newcommand\@fb@secFB{\FloatBarrier
    \gdef\@fb@afterHHook{\@fb@topbarrier \gdef\@fb@afterHHook{}}}%
  \g@addto@macro\@afterheading{\@fb@afterHHook}%
  \gdef\@fb@afterHHook{}%
}
\makeatother
\newcommand{\GeV}{\text{GeV}}
\newcommand{\MeV}{\text{MeV}}
\newcommand{\fm}{\text{fm}}
\newcommand{\tr}{\text{tr}\;}
\newcommand{\mydot}[1]{\tikz{\filldraw[draw=#1,fill=#1] (0,0) circle (3pt);}\ \ }
\newcommand{\mysquare}[1]{\tikz{\filldraw[draw=#1,fill=#1] (0,0) rectangle (6pt,6pt);}\ \ }
\newcommand{\mytriangle}[1]{\tikz{\filldraw[draw=#1,fill=#1] (-3pt,0) -- (3pt,0) -- (0pt,6pt);}\ \ }
\definecolor{H102}{HTML}{1F77B4}
\definecolor{N202}{HTML}{FF7F0e}
\definecolor{X250}{HTML}{2CA02C}
\definecolor{X251}{HTML}{D62728}
\definecolor{N200}{HTML}{9467BD}
\definecolor{N203}{HTML}{8C564B}
\definecolor{D201}{HTML}{17BECF}
\definecolor{N300}{HTML}{7F7F7F}
\definecolor{N302}{HTML}{7F7F7F}
\definecolor{B450}{HTML}{BCBD22}
\definecolor{J500}{HTML}{17BECF}

\makeatletter
\newcommand{\oset}[3][0ex]{%
  \mathrel{\mathop{#3}\limits^{
    \vbox to#1{\kern-2\ex@
    \hbox{$\scriptstyle#2$}\vss}}}}
\makeatother
\title{Hyperon couplings from $N_f = 2 + 1$ lattice QCD}
\author{
	Gunnar S.~Bali$^a$,
	Sara Collins$^a$,
	Piotr Korcyl$^{ab}$,
	Rudolf Rödl$^a$,
	\speaker{Simon Weishäupl}$^a$ and
	Thomas Wurm$^a$\\
    \llap{$^a$}Institut für Theoretische Physik, Universität Regensburg, 93040 Regensburg, Germany\\
    \llap{$^b$}Marian Smoluchowski Institute of Physics, Jagiellonian University, ul.\ {\L}ojasiewicza 11,\\30-348 Krak\'ow, Poland\\
    E-mail:
    \email{gunnar.bali@ur.de},
    \email{sara.collins@ur.de},
    \email{piotr.korcyl@uj.edu.pl},
    \email{rudolf.roedl@ur.de},
    \email{simon.weishaeupl@ur.de},
    \email{thomas.wurm@ur.de}
}

\abstract{
  We compute various (generalized) isovector charges of the octet baryons.
  These include $g_A$, $g_T$ and $g_S$ as well as the unpolarized,
  polarized and transversity parton distribution function (PDF)
  momentum fractions $\langle x\rangle_{u^+-d^+}$,
  $\langle x\rangle_{\Delta u^--\Delta d^-}$ and
  $\langle x\rangle_{\delta u^+-\delta d^+}$.
  The simulations are carried out on a subset of the (isospin symmetric)
  $N_f=2+1$ flavour Coordinated Lattice Simulations (CLS) gauge ensembles
  with lattice spacings ranging from $a\approx 0.086\,\fm$ down to
  $a\approx 0.050\,\fm$.
  First results on the breaking of flavour symmetry and the low energy
  constants $F$ and $D$ are presented. While SU(3) flavour symmetry
  violations are found to be sizeable for
  $g_A=\langle \mathds{1}\rangle_{\Delta u^+-\Delta d^+}$, these are quite small
  for $g_T=\langle \mathds{1}\rangle_{\delta u^--\delta d^-}$ and
  $\langle x\rangle_{u^+-d^+}$.
}
\FullConference{
    37th International Symposium on Lattice Field Theory -- Lattice2019\\
	16--22 June 2019\\
	Wuhan, China
}
\ShortTitle{Hyperon couplings from $N_f = 2 + 1$ lattice QCD}

\begin{document}
\section{Introduction}
In recent years quite a few lattice calculations of the nucleon axial
charge $g_A^N$ have been carried out, assuming isospin symmetry,
most of which are compatible with the
experimental ratio $g_A^N/g_V^N = 1.2732(23)$~\cite{PhysRevD.98.030001},
as measured from neutron $\beta$ decay. On the lattice
$g_A^N$ can be accessed, e.g., via the proton matrix element
$\langle p|\bar{u}\gamma_5\gamma_\mu u - \bar{d}\gamma_5\gamma_\mu d |p\rangle$.
The hyperon axial charges $g_A^B$ for $B\neq N$ are less well known since
a direct measurement, e.g., of $\Sigma^-\rightarrow\Sigma^0e^-\bar{\nu}_e$,
is unrealistic. So far these can only be inferred,
assuming an approximate flavour symmetry, from $\beta$ decays such as
$\Xi^-\rightarrow\Lambda\ell\bar{\nu}_{\ell}$ or
$\Sigma^-\rightarrow n\ell\bar{\nu}_{\ell}$,
where an $s$-quark is converted into a $u$-quark. In terms
of lattice simulations only few direct
results~\cite{Lin:2007ap,Erkol:2009ev,Gockeler:2011ze,Cooke:2012xv,Alexandrou:2016xok,Savanur:2018jrb} of these charges exist.
Such determinations, however, are interesting both in terms of testing
the extent of SU(3) flavour symmetry and to determine the
low energy constants (LECs) $F$ and $D$ from first principles, which
also appear in other SU(3) BChPT expansions, e.g., of octet
baryon self-energies.

In addition to the axial charges we also compute the hyperon isovector
charges in other channels, in particular the scalar charges $g_S^B$,
and the tensor charges $g_T^B$ as well as the
unpolarized, polarized and transversity Bjorken momentum fractions
$\langle x \rangle^B_{u^+-d^+}$,
$\langle x \rangle^B_{\Delta u^--\Delta d^-}$ and
$\langle x \rangle^B_{\delta u^+-\delta d^+}$, respectively, where
$q^{\pm}=q\pm\bar{q}$.
Much less is known for these (generalized) charges since these
are even more elusive to experimental measurement than $g_A^B$.
\section{CLS gauge ensembles used}
\TABULAR{cllrcccc}{
		\toprule
		Ensemble & $\beta$ & $ a [\fm] $ & $N_t\cdot N_s^3$ & $M_\pi [\MeV]$ & $LM_\pi$  & $t/a$                     & $N_{\text{configs}}$ \\
		\hline 
		\mydot{H102} H102     & 3.4     & $ 0.0856 $  & $96\cdot 32^3$  & 355          & 4.9         & [8(2), 10(2), 12(2), 14(2)]  & 2004 \\ 
		\hline
		\mytriangle{N202} N202     & 3.55    & $ 0.0642 $  & $128\cdot 48^3$ & 412          & 6.4         & [11, 14(2), 16(2), 19(4)]    & 1768 \\ 
		\mytriangle{X250} X250     &         &             & $64\cdot 48^3$  & 348         & 5.4         & [11, 14(2), 16(4), 19(4)]    & 345  \\ 
		\mytriangle{X251} X251     &         &             & $64\cdot 48^3$  & 269          & 4.2         & [11, 14(2), 16(4), 19(4)]    & 436  \\ 
		\mydot{N203} N203     &         &             & $128\cdot 48^3$ & 346          & 5.4         & [11(2), 14(2), 16(2), 19(2)] & 1543 \\ 
		\mydot{N200} N200     &         &             & $128\cdot 48^3$ & 284          & 4.4         & [11, 14, 16, 19]             & 1712 \\ 
		\mysquare{D201} D201     &         &             & $128\cdot 64^3$ & 199          & 4.1         & [11(2), 14(2), 16(2), 19(2)] & 1078 \\ 
		\hline 
		\mydot{N302} N302     & 3.7     & $ 0.0497 $  & $128\cdot 48^3$ & 347          & 4.1         & [14, 17, 21, 24]             & 1383 \\ 
		\bottomrule}
      {CLS gauge ensembles analysed here. $t$ denotes the source-sink
      separations and bracketed digits indicate the number of
      measurements carried out for each distance on each configuration.
    \label{tab:cls_ensembles}}
For our analysis we employ gauge ensembles generated by the
Coordinated Lattice Simulations~(CLS)~\cite{Bruno:2014jqa} effort,
that combine the $N_f = 2 + 1$ non-perturbatively
$\mathcal{O}(a)$ improved Wilson fermionic action with the tree-level
Symanzik improved gauge action. To avoid freezing of the topological charge
most of these ensembles have open boundary conditions in
time~\cite{Luscher:2011kk}. The ensembles are generated along
three different trajectories, namely:
\begin{itemize}[topsep=0pt,parsep=0pt,itemsep=0pt]
\item $\tr M = \text{const}$: keeping the trace of the quark mass matrix constant near its physical value~\cite{Bruno:2014jqa}, thereby increasing the
  strange quark mass while the light quark mass is decreased.
\item $\widehat{m}_s \approx \widehat{m}_s^{\text{ph}}$: setting the renormalized strange quark mass to its physical value~\cite{PhysRevD.94.074501}.
\item $m_\ell = m_s$: the symmetric line.
\end{itemize}
This enables us to extrapolate to the physical point along two different quark
mass trajectories
($\tr M \approx \tr M^{\text{ph}}$ and
$\widehat{m}_s\approx\widehat{m}_s^{\text{ph}}$),
while along the symmetric line an extrapolation to the $N_f=3$ chiral limit
can be carried out.

The parameters of the ensembles investigated so far
are listed in table~\ref{tab:cls_ensembles}. These cover a range of
pion masses from $\sim 410\,\MeV$ down to $\sim 200\,\MeV$, three
different lattice spacings $a$ and volumes with  $LM_\pi$ between 4.1 and 6.4.
D201 is on the $\widehat{m}_s\approx \widehat{m}_s^{\text{ph}}$
trajectory, which at this light pion mass is very close to the
$\tr M = \text{const}$ line. H102, N202, N203, N200 and N302 are on the
$\tr M = \text{const}$ line and
X250 and X251 (as well as N202) are on the SU(3) symmetric line.
\section{Definitions, numerical methods and the fit procedure}
The isovector charges are defined as matrix elements of local operators
at zero momentum transfer. Here we consider two kinds of (generalized) charges:
\begin{align}
    g_J^B &=
    \langle B | O(\Gamma_J) | B \rangle, \quad  J \in \{V,  A, T, S \},
    \\
	m_B \langle x \rangle^B_J 
	&=
	\langle B | O(\Gamma_J) | B \rangle, \quad J \in \{ u^+-d^+, \Delta u^- - \Delta d^-, \delta u^+ - \delta d^+ \},
\end{align}
where the $\Gamma$-structures of the latter currents contain one derivative and
in both cases we use isovector combinations
$O(\Gamma_J) = \bar{u}\Gamma_J u - \bar{d}\Gamma_J d$.
We destroy the nucleon, the $\Sigma$ and the $\Xi$ octet baryon
components with Dirac index $\alpha$,
using the interpolators
\begin{align}
    N_{\alpha} =p_\alpha= 
    \epsilon^{ijk} u^i_{\alpha}\left( u^{j\intercal} C\gamma_5 d^k \right),\quad
    \Sigma^+_{\alpha} =
    \epsilon^{ijk} u^i_{\alpha}\left( u^{j\intercal} C\gamma_5 s^k \right),\quad
    \Xi^0_{\alpha} =
    \epsilon^{ijk} s^i_{\alpha}\left( s^{j\intercal} C\gamma_5 u^k \right),
\end{align}
where $q\in\{u,d,s\}$ are Wuppertal smeared quark fields with
$m_s\geq m_{\ell}=m_u=m_d$. When constructing zero-momentum
two-point functions $C_{2pt,B}$, we project with
$\mathbb{P}_+=\tfrac12(\mathds{1}+\gamma_4)$ onto positive parity,
while for the three-point functions, we also project on helicity
differences if this is required.

The matrix elements are obtained from ratios of three-point over two-point
functions
\begin{align}
  \label{eq:rat}
    R_J^B(t,\tau) 
    &= \frac{C_{3pt,B}(t,\tau,\mathbf{q}=\mathbf{p}'=\mathbf{0},\Gamma_J)}{C_{2pt,B}(t,\mathbf{p}=\mathbf{0})} 
    \xlongrightarrow{t\gg\tau\to \infty} 
    \langle B | O(\Gamma_J) | B \rangle^{\text{latt}},
\end{align}
where $t$ is the temporal source-sink separation and $\tau$ the distance of
the current from the source.
In our case the discretization effects are of orders $a^2$
and $a$ for matrix elements without
and with derivatives, respectively. To quote results in the continuum
$\overline{\text{MS}}$ scheme at the scale $\mu=2\,\GeV$, we
renormalize the lattice matrix elements, multiplying these by
\begin{align}
	Z_J(g^2,a\mu) \left(1 + am_{\ell} b_J(g^2) + 3a \overline{m} \tilde{b}_J(g^2)\right),
\end{align}
where $3\overline{m}=\tr M$.
We non-perturbatively determine the renormalization with respect to the
intermediate RI'-SMOM scheme and employ the improvement coefficients $b_J(g^2)$
of~\cite{Korcyl:2016cmx}, where first estimates of $\tilde{b}_J$
were found to be compatible with zero.

With the exception of the $m_{\ell}=m_s$ ensembles
the three-point correlation functions are computed using the stochastic method
described in~\cite{Bali:2017mft} (see also~\cite{Evans:2010tg,Bali:2013gxx,Alexandrou:2013xon,Yang:2015zja}),
estimating a timeslice-to-all propagator (depicted as a wiggly line in
figure~\ref{fig:sct_diagram_color_coded}).
This allows us to factorize the three-point correlation function into two
parts, the spectator part~$S$ and the insertion part~$I$, which can be
evaluated separately, where we leave all Dirac indices (Greek letters)
as well as the stochastic and colour indices ($n$ and $k$) open:
\begin{align}
 &C(
 \mathbf{p}^\prime,
 \mathbf{q},
 x^\prime_4,
 y_4,
 x_4
 )^{
 {\color{red}
 \alpha^\prime\,
 \alpha\,
 \beta^\prime\,
 \beta\,
 \delta^\prime}
 {\color{blue}
 \delta\,
 \gamma^\prime
 \gamma}
 }_{
 {\color{red}
 U \,
 D \,}
 \color{blue}{
 U \,
 U \,}
 }
 =
 \frac{1}{N_{\mathrm{sto}}}\sum  \limits_{n=1}^{N_{\mathrm{sto} }}
 \!
 \sum \limits_{k=1}^3
 \left(
 S_{
 	\color{red}{U
 	D}
 }(
 \mathbf{p}^\prime,
 x^\prime_4,
 x_4
 )^{
 {\color{red}
 \alpha^\prime\,
 \alpha\,
 \beta^\prime\,
 \beta\,
 \delta^\prime}
 }_{
 n
 k
 }
 \cdot\,
 I_{\color{blue}UU}(\mathbf{q}, y_4, x_4 )^{
 {\color{blue}
 \delta\,
 \gamma^\prime
 \gamma}
 }_{
 n
 k
 }
 \right).
 \label{eq:stoch_corr}
\end{align}
\FIGURE[r]{
	\includegraphics[width=.5\textwidth]{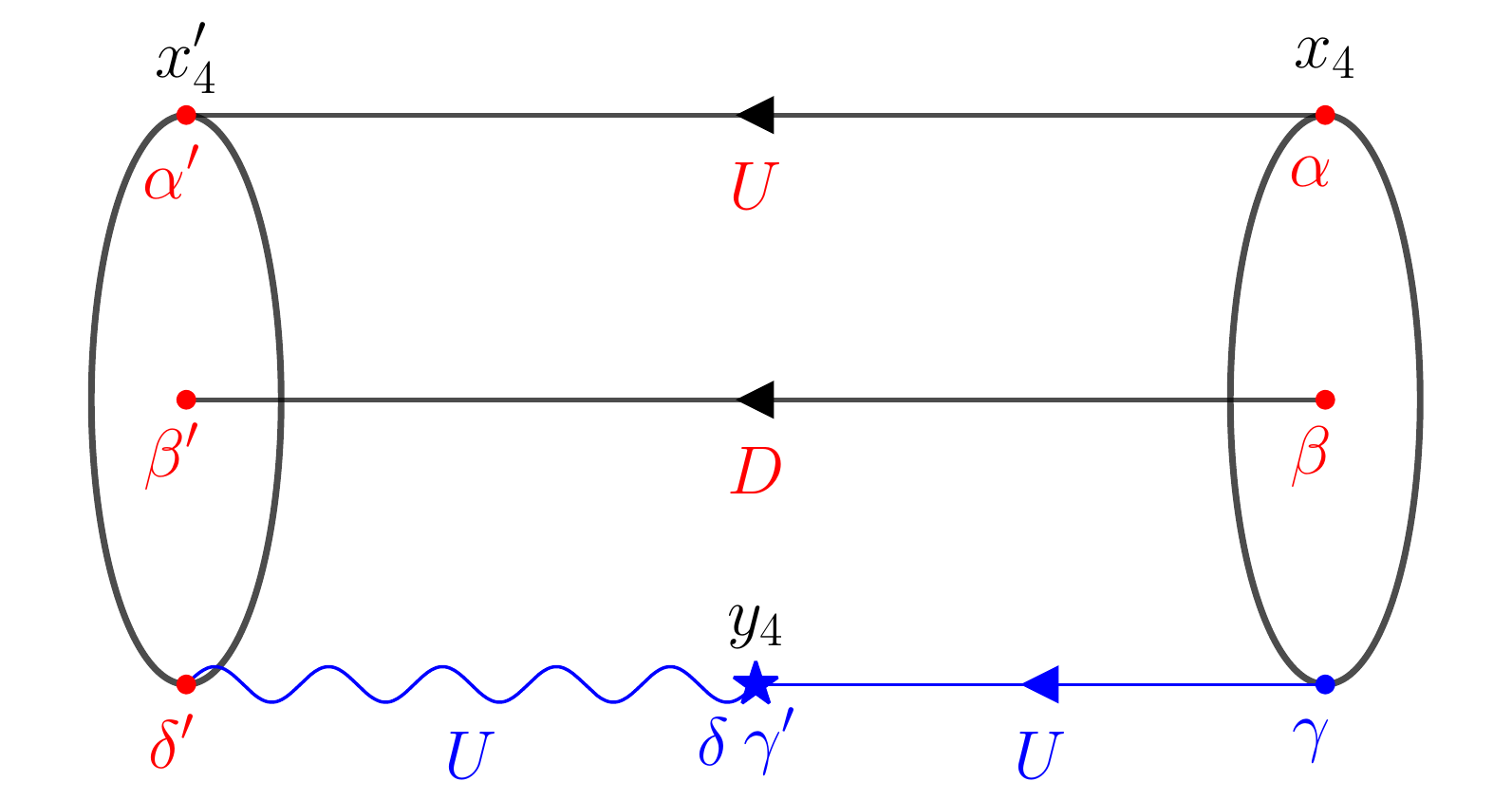}
	\caption{The computation of the stochastic three-point function. The red and blue indices correspond to the spectator and insertion part, see eq.~\eqref{eq:stoch_corr}. The blue wiggly line represents the stochastic propagator, whereas solid lines are standard point-to-all propagators.}
	\label{fig:sct_diagram_color_coded}
}
The baryon source time $x_4$ and the current insertion time $y_4$, where
the spatial momentum $\mathbf{q}$ is injected, are varied, while
the sink time $x_4'$ of the baryon with momentum $\mathbf{p}'$,
where the stochastic source is located, is fixed. We have a similar backward
sink (not shown) to enable forward-backward time averaging.
For the baryons other than the proton, we substitute the\break quark flavours
accordingly. We then compute the contractions for the matrix ele-\break
ments
of interest. In this analysis
we restrict ourselves to non-flavour changing currents and
$\mathbf{p}'=\mathbf{q}=\mathbf{0}$.

Note that for different baryon sinks or momenta no additional
inversions are needed. However, stochastic noise is added
to the gauge noise. This is eventually offset
by averaging over three-point
functions in the forward and backward directions as well as over
equivalent polarizations and momentum combinations
at very little computational overhead. Any baryonic three-point function with
currents containing (in our case) up to one derivative can be
computed at the analysis stage from the building blocks
$S$ and $I$ that we store, making this method extremely versatile.

In addition to statistical errors there  will be systematic uncertainties
related to the precision of the renormalization constants and improvement
coefficients, the continuum limit extrapolation, finite volume effects and
the use of unphysical quark masses. Not all of these systematics can be
explored on the presently analysed subset of CLS ensembles. With all
lattice extents $L>4.1/M_\pi$, finite size effects should be negligible.
Before discussing the quark mass dependence, we will study the effect of
excited state contributions.
\FIGURE{
	\includegraphics[width=.85\textwidth]{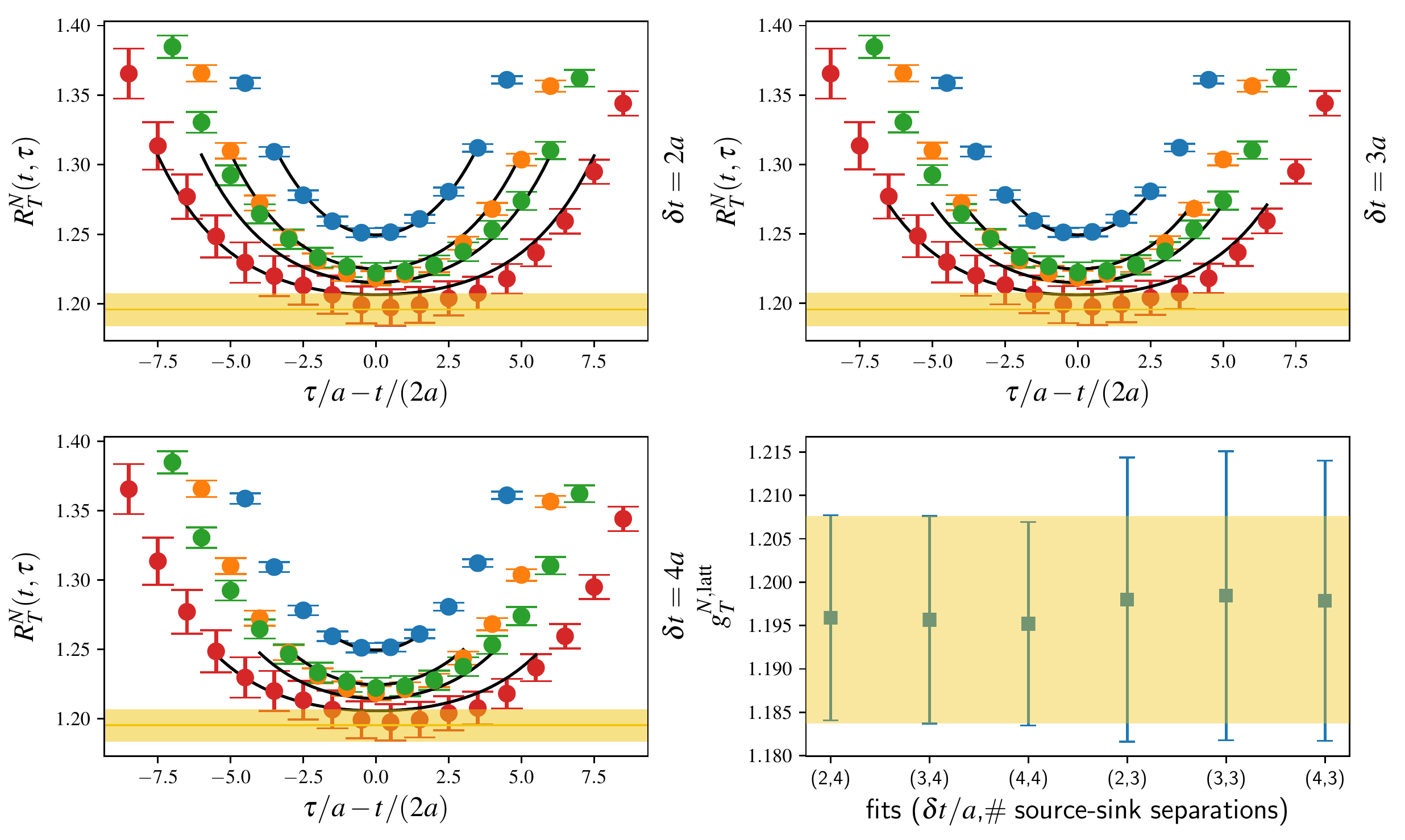}
	\caption{Ratios $R_T^N$ (see eq.~\eqref{eq:rat}) for the
          tensor charge of the nucleon $g_T^N$ on ensemble N203
          for different source-sink separations $t$, together
          with the fits~\eqref{eq:fit} covering different intervals
          $\tau\in[\delta t,t-\delta t]$.
          We indicate the results of these and three additional fits
          where we omit the smallest $t$ value in the bottom right panel.}
	\label{fig:N203_nucleon_gt}\vspace*{-.4cm}}
For all ensembles we use a set of four different, approximately matched
source-sink separations
$t/\fm=(x_4'-x_4)/\fm\oset[-.3ex]{\sim}\in\{0.7, 0.8, 1.0, 1.2\}$.
In figure~\ref{fig:N203_nucleon_gt} we show the ratio~$R_T^N$  (see eq.~\eqref{eq:rat}) for the
tensor channel, in which excited state contributions are clearly visible.
To obtain the final result, we carry out simultaneous fits to the
two-point function and the ratios using all four (or three) source-sink
separations, according to the ansatz
\begin{align}\label{eq:fit}
	C_{2pt,B}(t) =  A_0 e^{-m_B t} \left( 1 + A_1 e^{-\Delta m_B t} \right),\quad
	R_J^B(t,\tau)  =  B_0 + B_1 e^{-\Delta m_B t / 2} \cosh (\Delta m_B (\tau - t/2) ) + B_2 e^{-\Delta m_B t}
\end{align}
with the mass difference $\Delta m_B = m_B^\prime - m_B$. 
In these fits the position of the current
$\tau=y_4-x_4$ is allowed to run within the
interval $\tau\in[\delta t,t-\delta t]$, where
$\delta t\in\{2a, 3a, 4a\}$. Overall, we find stable results
for all six fits as can be seen in the bottom right panel
of the figure, where the error band indicates our final result.

\section{Results and Outlook}
\FIGURE{
	\includegraphics[width=.32\textwidth]{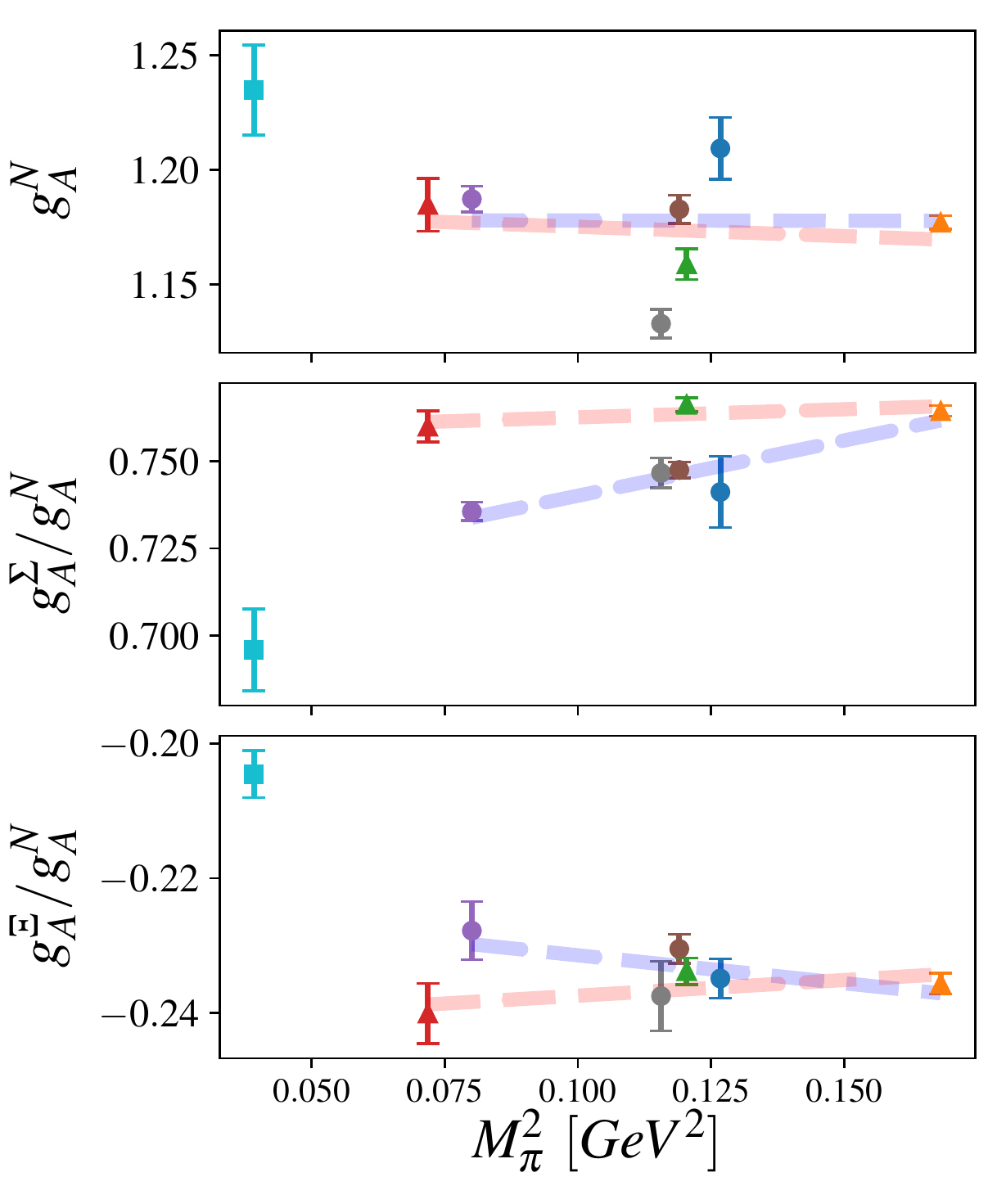}
	\includegraphics[width=.32\textwidth]{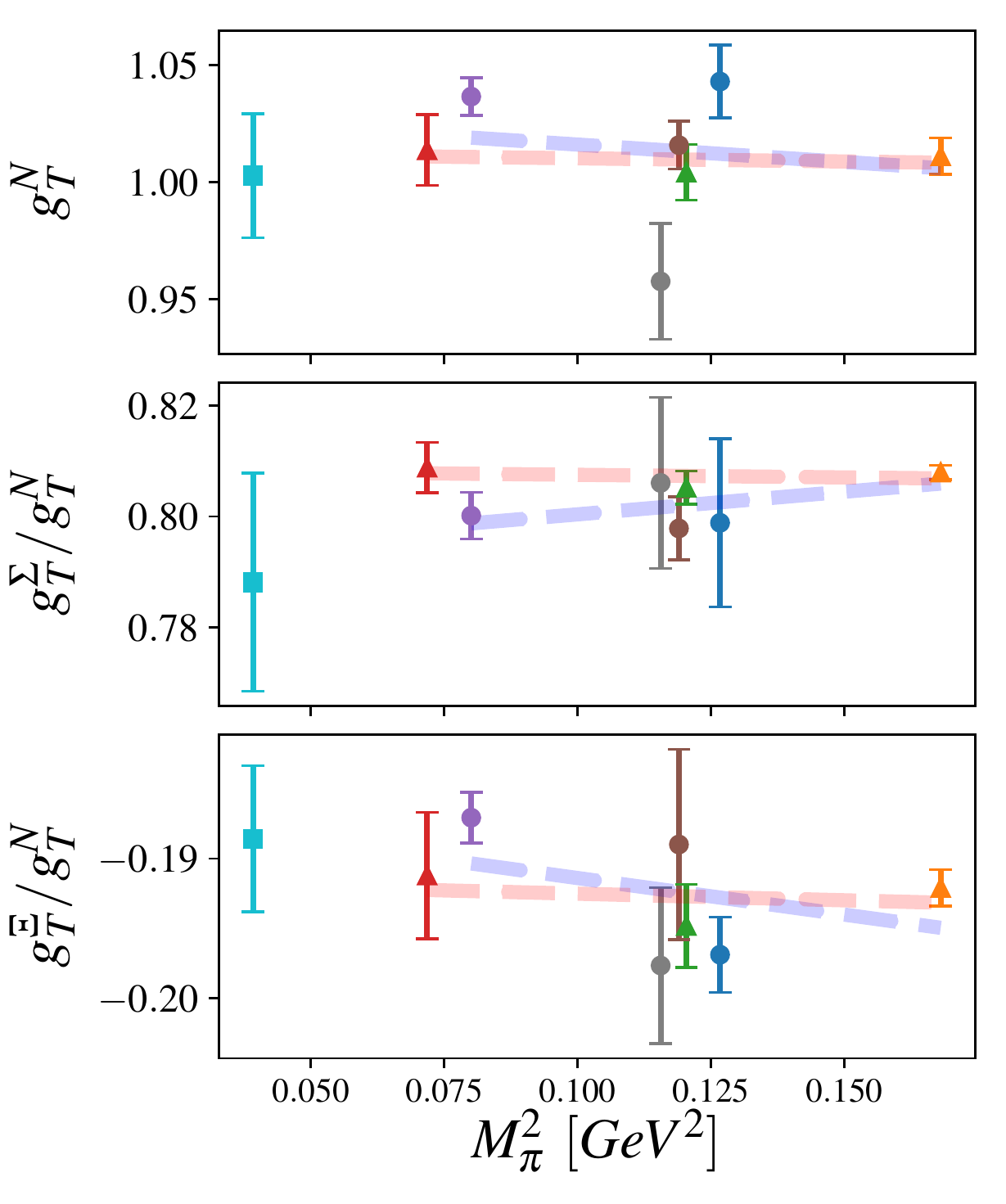}
	\includegraphics[width=.33\textwidth]{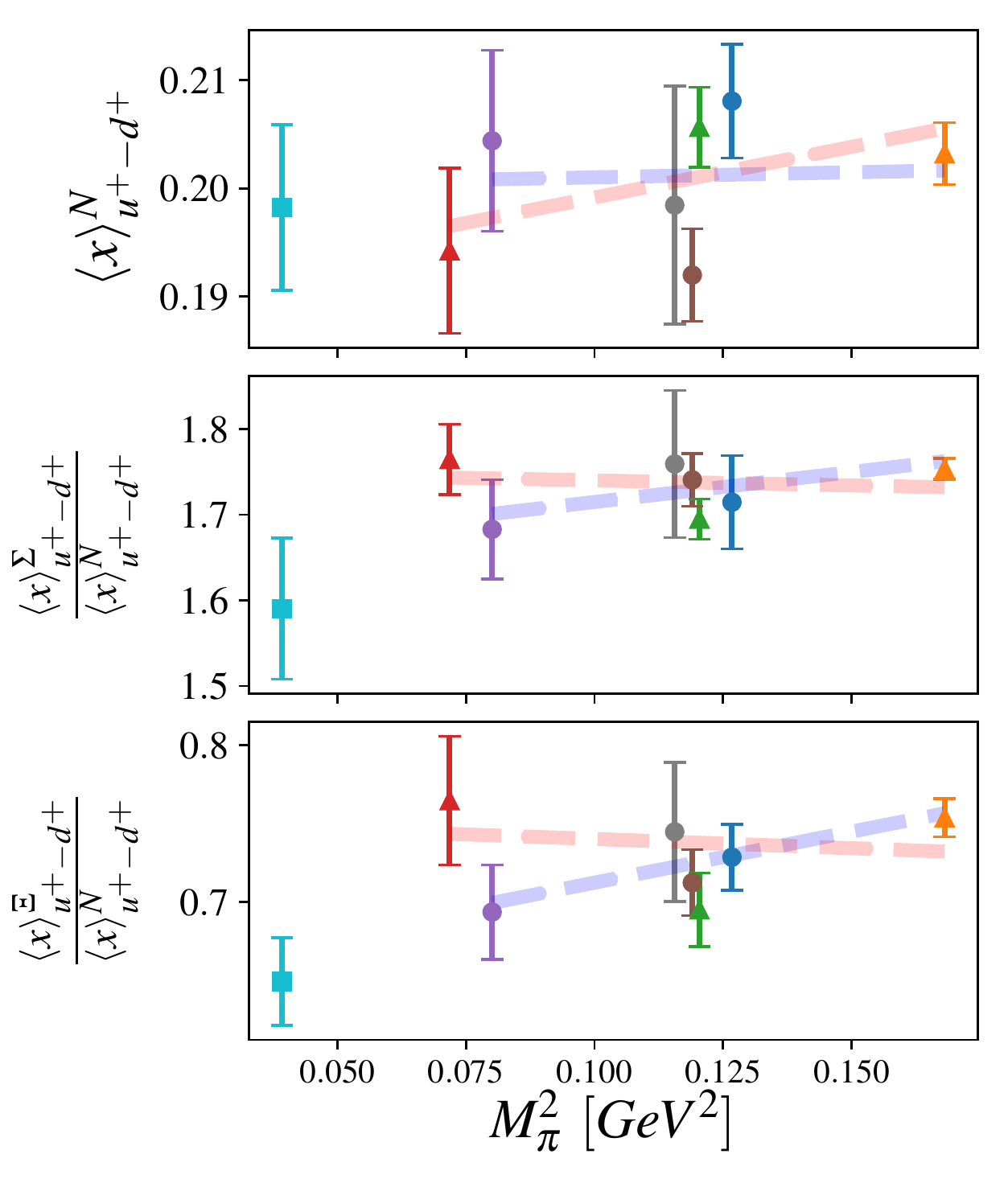}
	\caption{Axial and tensor charges $g_A^B$, $g_T^B$ as well as $\langle x \rangle^B_{u^+-d^+}$, for $B\in\{\Sigma, \Xi\}$ divided by the nucleon's.}
	\label{fig:hyperon_ga_extrapol}\vspace*{-.2cm}}
In figure~\ref{fig:hyperon_ga_extrapol} we show results for the
examples of the axial and tensor charges and the unpolarized isovector
PDF Mellin moment $\langle x\rangle^B_{u^+-d^+}$. The dashed red lines that
are drawn to guide the eye connect
the $m_\ell=m_s$ points that will approach the SU(3) chiral limit while
the blue lines connect the $\tr M\approx\tr M^{\text{ph}}$ points. The left-most
point corresponds to D201, which is on the
$\widehat{m}_s\approx\widehat{m}_s^{\text{ph}}$ line, but also near the
$\tr M=\tr M^{\text{ph}}$ trajectory, due to the small
pion mass. The first row shows the nucleon charges which are
subject to visible cut-off effects. When normalizing hyperon charges
with respect to the nucleon charges (second and third rows)
the renormalization and improvement factors cancel and
the data collapse onto two curves for the two quark
mass trajectories, indicating that lattice spacing
effects cancel to a large extent.

\FIGURE[t]{
	\includegraphics[width=.85\textwidth]{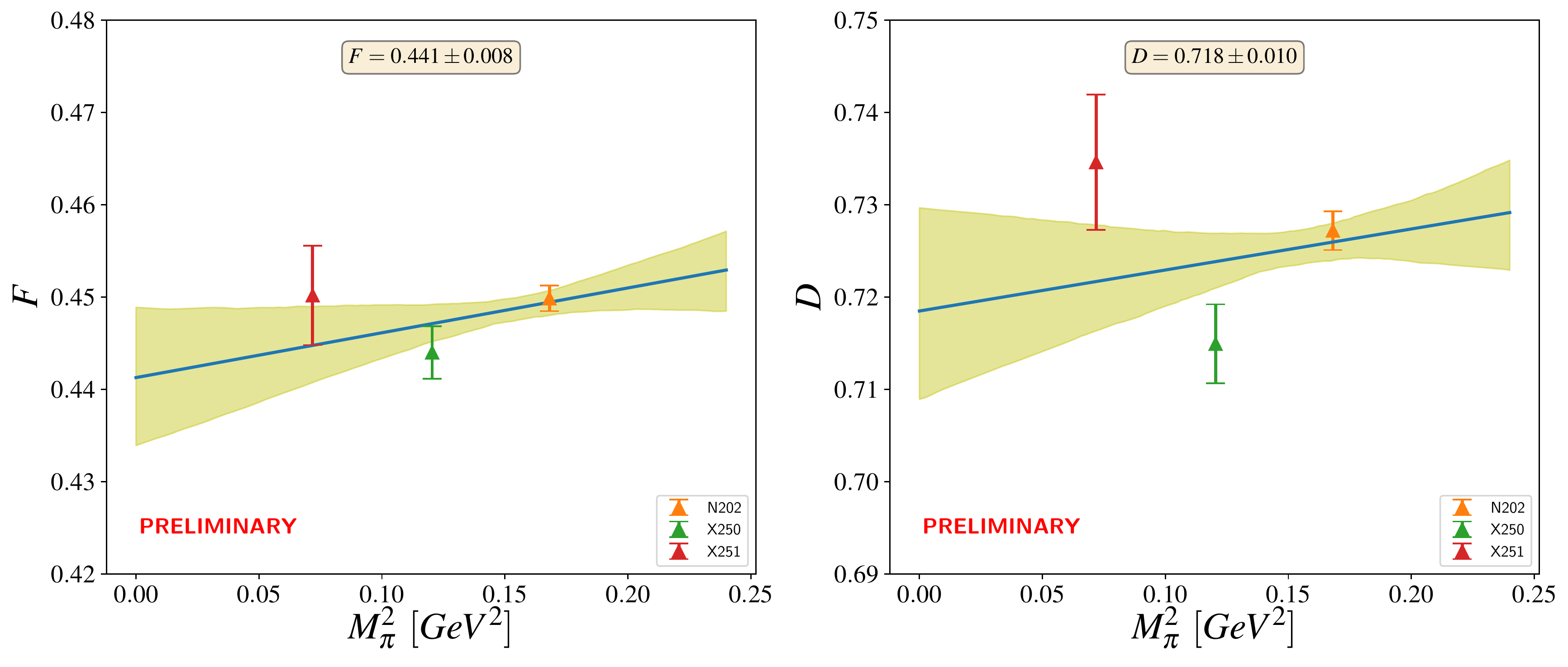}
	\caption{Linear extrapolation in $M_\pi^2$ of
          the LECs $F$ and $D$ for the ensembles on the symmetric line.}
	\label{fig:f_d_ga_extrapol}\vspace*{-.2cm}}
\FIGURE[h]{
	\includegraphics[width=.85\textwidth]{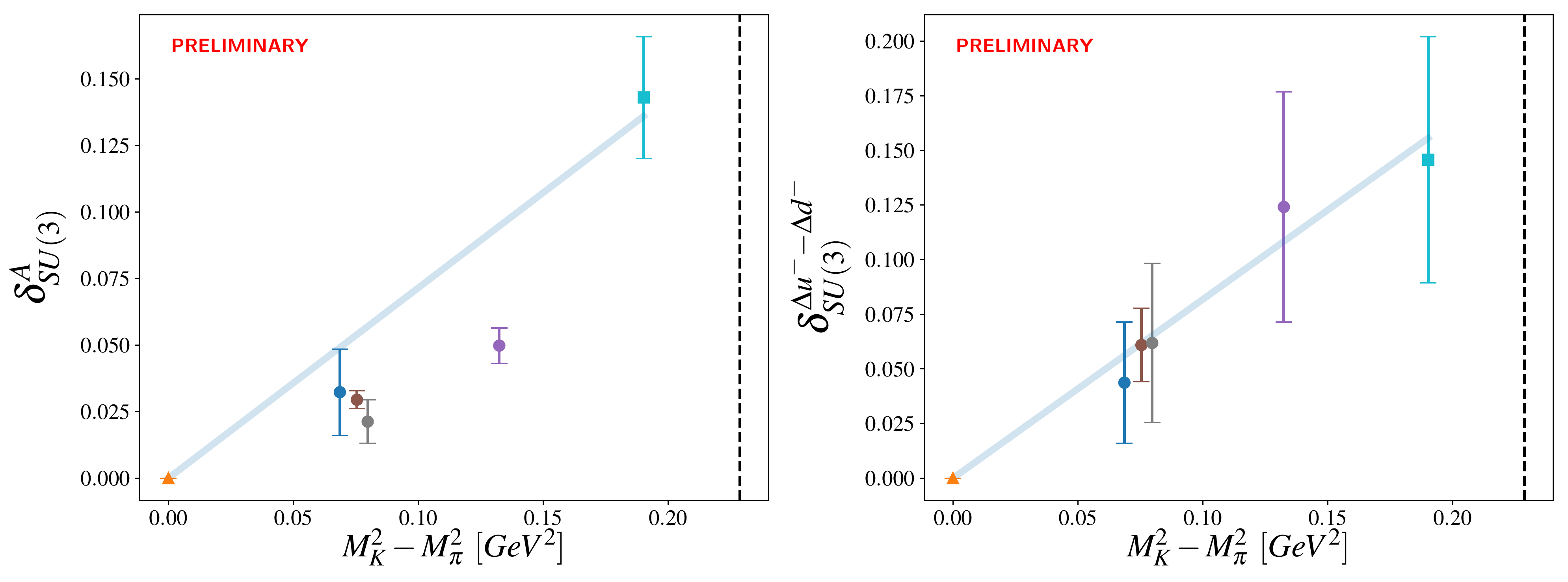}
	\caption{Symmetry breaking parameter (see eq.~\eqref{eq:sym_breaking})
          for $g_A$ and
          $\langle x \rangle_{\Delta u^- - \Delta d^-}$, where $M_K^2 - M_\pi^2 \propto
          m_s - m_{\ell}$. The dashed vertical line indicates the
          physical point. The blue lines are only drawn to guide the eye.}
	\label{fig:symmetry_breaking_axial_charges}\vspace*{-.1cm}}
Assuming SU(3) flavour symmetry, the axial vector charges can be expressed
as combinations of the two LECs $F$ and $D$:
\begin{align}
	g_A^N = F + D,\qquad
	g_A^\Sigma = 2F,\qquad
	g_A^\Xi = F - D.
        \label{eq:su3}
\end{align}
We can directly extract these LECs, extrapolating the combinations
$F=(g_A^N+g_A^\Xi)/2$ and
$D=(g_A^N-g_A^\Xi)/2$ along the $m_{\ell}=m_s$ trajectory as
linear functions in $M_{\pi}^2$, see
figure~\ref{fig:f_d_ga_extrapol}. Our values are consistent
with the recent lattice results obtained in~\cite{Savanur:2018jrb},
albeit the latter correspond to the SU(2) and not to the SU(3) chiral
limit.

Finally, we investigate SU(3) flavour symmetry breaking effects.
Note that according to eq.~\eqref{eq:su3}
$(g^N_J+g^\Xi_J)/g^{\Sigma}_J=(2F_J)/(2F_J)=1$ holds for $m_{\ell}=m_s$.
Thus the breaking effect for a current $J$ can, e.g., be quantified
in terms of the parameter
\begin{align}
 	\delta^J_{\text{SU(3)}}=\left(g^N_J+g^\Xi_J\right)/g^{\Sigma}_J-1.
 	\label{eq:sym_breaking}
\end{align}
These effects are small for $g_T$ and $\langle x \rangle_{u^+ - d^+}$
as is also evident from figure~\ref{fig:hyperon_ga_extrapol}, where
the blue and red data sets do not significantly deviate from each other.
However, for the axial moments
$g_A=\langle\mathds{1}\rangle_{\Delta u^+ - \Delta d^+}$ and
$\langle x \rangle_{\Delta u^- - \Delta d^-}$ we find symmetry
breaking effects of about 10\% at physical quark masses, see
figure~\ref{fig:symmetry_breaking_axial_charges}.
This will be investigated in detail, employing the systematic approach
used by QCDSF~\cite{Cooke:2012xv}.

We are in the process of analysing additional ensembles that should enable
us to carry out controlled physical point and continuum limit extrapolations.
We also plan to determine the $\Delta S=1$ (generalized) form factors
to complete the picture of flavour symmetry violation and to determine
the relevant baryon ChPT LECs.

\noindent {\bf Acknowledgements}:
  The work of GB and SW is funded by the
  German BMBF grant 05P18WRFP1.
  Partial supported was also granted by the EU ITN EuroPLEx (grant 813942)
  and by the German DFG (SFB/TRR 55). We thank
  Benjamin Gl\"a\ss{}le, Simon Heybrock and Marius L\"offler for
  co-developing some of the software used here.
  We gratefully acknowledge computing time granted by the
  John von Neumann Institute for Computing (NIC), provided on the Booster
  partition of the supercomputer JURECA~\cite{jureca} at
  \href{http://www.fz-juelich.de/ias/jsc/}{J\"ulich Supercomputing
    Centre (JSC)}. Additional simulations were carried out at the
  QPACE~3 Xeon Phi cluster of SFB/TRR~55.\\[-.8cm]

\providecommand{\href}[2]{#2}\begingroup\raggedright\endgroup

\begin{thebibliography}{10}
\setlength{\parskip}{0pt}
\setlength{\itemsep}{0pt plus 0.3ex}
\vspace*{-.1cm}
\bibitem{PhysRevD.98.030001}
{\scshape Particle Data Group}: M.~Tanabashi et~al., \emph{Review of particle
  physics}, \href{https://doi.org/10.1103/PhysRevD.98.030001}{\emph{Phys. Rev.
  D} {\bfseries 98} (2018) 030001}.
\bibitem{Lin:2007ap}
H.-W. Lin and K.~Orginos, \emph{{First calculation of hyperon axial couplings
  from Lattice QCD}},
  \href{https://doi.org/10.1103/PhysRevD.79.034507}{\emph{Phys. Rev. D}
  {\bfseries 79} (2009) 034507}
  [\href{https://arxiv.org/abs/0712.1214}{{\ttfamily 0712.1214}}].
\bibitem{Erkol:2009ev}
G.~Erkol, M.~Oka and T.~T. Takahashi, \emph{{Axial charges of octet baryons in
  two-flavor Lattice QCD}},
  \href{https://doi.org/10.1016/j.physletb.2010.02.016}{\emph{Phys. Lett. B}
  {\bfseries 686} (2010) 36} [\href{https://arxiv.org/abs/0911.2447}{{\ttfamily
  0911.2447}}].
\bibitem{Gockeler:2011ze}
{\scshape QCDSF/UKQCD} collaboration: M.~G\"ockeler et~al., \emph{{Baryon axial charges and momentum
  fractions with $N_f=2+1$ dynamical fermions}},
  \href{https://doi.org/10.22323/1.105.0163}{\emph{PoS} {\bfseries LATTICE2010}
  (2010) 163} [\href{https://arxiv.org/abs/1102.3407}{{\ttfamily 1102.3407}}].
\bibitem{Cooke:2012xv}
{\scshape QCDSF/UKQCD} collaboration: A.~N.~Cooke et~al., \emph{{The effects of flavour symmetry breaking on hadron matrix
  elements}}, \href{https://doi.org/10.22323/1.164.0116}{\emph{PoS} {\bfseries
  LATTICE2012} (2012) 116} [\href{https://arxiv.org/abs/1212.2564}{{\ttfamily
  1212.2564}}].
\bibitem{Alexandrou:2016xok}
C.~Alexandrou, K.~Hadjiyiannakou and C.~Kallidonis, \emph{{Axial charges of
  hyperons and charmed baryons using $N_f=2+1+1$ twisted mass fermions}},
  \href{https://doi.org/10.1103/PhysRevD.94.034502}{\emph{Phys. Rev. D}
  {\bfseries 94} (2016) 034502}
  [\href{https://arxiv.org/abs/1606.01650}{{\ttfamily 1606.01650}}].
\bibitem{Savanur:2018jrb}
A.~Savanur and H.-W. Lin, \emph{{Lattice-QCD determination of the hyperon axial
  couplings in the continuum limit}},
  \href{https://arxiv.org/abs/1901.00018}{{\ttfamily 1901.00018}}.
\bibitem{Bruno:2014jqa}
{\scshape CLS}: M.~Bruno et~al., \emph{{Simulation of QCD with $N_{f} = 2 + 1$ flavors of
  non-perturbatively improved Wilson fermions}},
  \href{https://doi.org/10.1007/JHEP02(2015)043}{\emph{JHEP} {\bfseries 02}
  (2015) 043} [\href{https://arxiv.org/abs/1411.3982}{{\ttfamily 1411.3982}}].
\bibitem{Luscher:2011kk}
M.~L{\"u}scher and S.~Schaefer, \emph{{Lattice QCD without topology barriers}},
  \href{https://doi.org/10.1007/JHEP07(2011)036}{\emph{JHEP} {\bfseries 07}
  (2011) 036} [\href{https://arxiv.org/abs/1105.4749}{{\ttfamily 1105.4749}}].
\bibitem{PhysRevD.94.074501}
  {\scshape RQCD} collaboration: G.~S.~Bali, E.~E.~Scholz, J.~Simeth and W.~S\"oldner,
  \emph{Lattice simulations with
  $n_f=2+1$ improved wilson fermions at a fixed strange quark mass},
  \href{https://doi.org/10.1103/PhysRevD.94.074501}{\emph{Phys. Rev. D}
  {\bfseries 94} (2016) 074501}.
\bibitem{Korcyl:2016cmx}
P.~Korcyl and G.~S.~Bali, \emph{{Non-perturbative determination of improvement
  coefficients using coordinate space correlators in $N_f=2+1$ lattice QCD}},
  \href{https://doi.org/10.22323/1.256.0190}{\emph{PoS} {\bfseries LATTICE2016}
  (2016) 190} [\href{https://arxiv.org/abs/1609.09477}{{\ttfamily
  1609.09477}}].
\bibitem{Bali:2017mft}
   {\scshape RQCD} collaboration: M.~L{\"o}ffler et~al., \emph{{Baryonic and mesonic 3-point functions with
  open spin indices}},
  \href{https://doi.org/10.1051/epjconf/201817506014}{\emph{EPJ Web Conf.}
  {\bfseries 175} (2018) 06014}
  [\href{https://arxiv.org/abs/1711.02384}{{\ttfamily 1711.02384}}].
\bibitem{Evans:2010tg}
R.~Evans, G.~Bali and S.~Collins, \emph{{Improved semileptonic form factor
  calculations in Lattice QCD}},
  \href{https://doi.org/10.1103/PhysRevD.82.094501}{\emph{Phys. Rev. D}
  {\bfseries 82} (2010) 094501}
  [\href{https://arxiv.org/abs/1008.3293}{{\ttfamily 1008.3293}}].
\bibitem{Bali:2013gxx}
{\scshape RQCD} collaboration: G.~S. Bali et~al., \emph{{Nucleon structure from stochastic estimators}},
  \href{https://doi.org/10.22323/1.187.0271}{\emph{PoS} {\bfseries LATTICE2013}
  (2014) 271} [\href{https://arxiv.org/abs/1311.1718}{{\ttfamily 1311.1718}}].
\bibitem{Alexandrou:2013xon}
{\scshape ETM} collaboration: C.~Alexandrou et~al., \emph{{A stochastic method for computing hadronic
  matrix elements}},
  \href{https://doi.org/10.1140/epjc/s10052-013-2692-3}{\emph{Eur. Phys. J. C}
  {\bfseries 74} (2014) 2692}
  [\href{https://arxiv.org/abs/1302.2608}{{\ttfamily 1302.2608}}].
\bibitem{Yang:2015zja}
Y.-B. Yang, A.~Alexandru, T.~Draper, M.~Gong and K.-F. Liu, \emph{{Stochastic
  method with low mode substitution for nucleon isovector matrix elements}},
  \href{https://doi.org/10.1103/PhysRevD.93.034503}{\emph{Phys. Rev. D}
  {\bfseries 93} (2016) 034503}
  [\href{https://arxiv.org/abs/1509.04616}{{\ttfamily 1509.04616}}].
\bibitem{jureca}
  {J\"{u}lich Supercomputing Centre: D.~Krause and P.~Th\"ornig},
  \emph{{JURECA: Modular supercomputer at
  J\"{u}lich Supercomputing Centre}},
  \href{https://doi.org/10.17815/jlsrf-4-121-1}{\emph{Journal of large-scale
  research facilities} {\bfseries 4} (2018) A132}.
\end{thebibliography}
\end{document}